# A new high-throughput method using additive manufacturing for alloy design and heat treatment optimization

Yunhao Zhao, Noah Sargent, Kun Li, Wei Xiong*
Physical Metallurgy and Materials Design Laboratory,
Department of Mechanical Engineering and Materials Science,
University of Pittsburgh, Pittsburgh, PA 15261, USA
* Corresponding Author, Email: weixiong@pitt.edu, w-xiong@outlook.com
URL: https://www.pitt.edu/~weixiong
Tel. +1 (412) 383-8092, Fax: +1 (412) 624-4846

## Abstract

Many alloys made by Additive Manufacturing (AM) require careful design of post-heat treatment as an indispensable step of microstructure engineering to further enhance the performance. We developed a high-throughput approach by fabricating a long-bar sample heat-treated under a monitored gradient temperature zone for phase transformation study to accelerate the post-heat treatment development of AM alloys. This approach has been proven efficient in determining the aging temperature with peak hardness. We observed that the precipitation strengthening is predominant for the studied superalloy by laser powder bed fusion, and the grain size variation is insensitive on temperature between 605 and 825°C. This new approach can be applied to post-heat treatment optimization of other materials made by AM, and further assist new alloy development.

Composition design and post-heat treatment optimization are the two main aspects of materials development [1,2]. Materials Genome highlights the importance of the rapid generation of the database to accelerate the materials development [1,3]. Therefore, high-throughput (HT) modeling and experimentation are critical to design efficiency [4]. In additive manufacturing (AM), the feedstock materials design becomes more important since more processing parameters can directly impact the microstructure-property relationships of the as-fabricated materials. For example, the Inconel 718 alloy made by laser powder bed fusion (LPBF) process often exhibits large columnar grains with texture induced anisotropic properties [5,6]. Therefore, post-heat treatment is often desired to further improve materials performance through microstructure engineering. Particularly, the post-heat treatment is critical for most of the metallic materials by AM, because work hardening can no longer become an option for microstructure engineering. In this work, we couple the LPBF technique with the gradient temperature heat treatment (GTHT) process [2,7] as an effective HT tool to accelerate the post-heat treatment design for AM components.

In order to evaluate the proof of concept, we choose Inconel 718 superalloy as the studying alloy, since it is widely manufactured using LPBF, and all of the phases are known and easy to confirm. Inconel 718 is a Ni-based superalloy with exceptional high-temperature mechanical properties through precipitation strengthening [8–11]. The strengthening effect of Inconel 718 mainly derives from the major strengthening phase γ″ (Ni3Nb, bct_D0$_{22}$) and the minor strengthening phase γ′ (Ni$_3$(Al, Ti, Nb), fcc_L1$_2$). The two strengthening phases usually form during the aging processes [12,13]. Additionally, needle-shaped detrimental phase δ ($Ni_3Nb$, $D0_a$) may also precipitate along the grain or twin boundaries during the aging [12]. Therefore, it is important to perform a careful optimization of the aging conditions in order to improve the mechanical performance of Inconel 718. The traditional two-step aging method for Inconel 718 is at 720°C for 8 hours plus 620°C for 8 hours [14]; however, other aging temperatures are also used to achieve specific mechanical properties. For instance, aging at 790°C was adopted to coarsen γ″, resulting in improved toughness properties for oil-grade Inconel 718 [9,15,16]. A general study was conducted by Slama et al. [17] to investigate the effect of aging temperature on the microstructure and hardness evolution in hot-rolled Inconel 718; it was found that the aging at 680°C for 50 hours could produce the highest hardness of 500 $HV_{0.5}$. Despite the mentioned progress, the influence of the aging temperature on the AM Inconel 718 alloys remains unclear.

In this work, we fabricated a long-bar sample using the LPBF technique on the EOS M 290 machine. The printing process uses the laser power of 285 W, the scan velocity of 960 mm/s, the hatching space of 0.11 mm, and the hatching lines between the adjacent layers rotating 67°, are used for such a build. As shown in Fig. 1(a), a cuboid Inconel 718 build (with a composition of Ni-18.26Fe-18.87Cr-4.97Nb-2.97Mo-0.94Ti-0.46Al-0.03C-0.06Mn-0.23Co-0.05Cu-0.06Si, in wt.%) was printed. The Inconel 718 build was designed with twenty-three evenly distributed holes. These holes provided flexibility when choosing monitoring locations. In addition, the holes increased the surface area of the sample and further improved the convection heat transfer, which reduced the variation in the sample temperature relative to the air temperature. As a result, the air temperature calibration became more representative of the real sample temperature, which allowed the preemptive selection of the monitoring locations in the sample according to the actual needs. The flexibility of the setup of the high-throughput experiment was increased by adopting additive manufacturing methods for sample fabrication. Using this methodology, the current work significantly reduced the total time needed for heat treatment.

The long bar build was firstly encapsulated in a quartz tube with a backfilled Ar protective atmosphere and then homogenized at 1180°C for 1 hour to eliminate the microsegregation and the grain texture from the AM process. The microstructures of the as-built sample and the sample homogenized at 1180°C for 1 hour were reported in [18]. Afterward, the sample was quenched into ice-water. Before conducting the HT aging heat treatment, eight K-type thermocouples were fixed into eight equidistantly distributed holes using conductive high-temperature cement, as shown in Fig. 1(b). The thermocouples were connected to a computer via a data acquisition system to record the aging temperatures at each location throughout the aging process. The uncertainty of the temperature control is ±0.75%. The aging heat treatment was then carried out in a tube furnace with one end open to introduce gradient temperatures at different locations in the sample, as illustrated in Fig. 1(c). The furnace settings necessary to achieve the desired temperature gradient were calibrated by measuring the temperature as a function of position with a K-type thermocouple. Calibration of the temperature profile is kind of time-consuming as several attempts may be needed to find a good combination of the furnace settings and the plug location to obtain the desired temperature gradient. Thus, the HT experimental method described in this work mainly saves time for long term heat treatments. It is important to note that the achievable temperature gradient will change depending on the furnace. Therefore, if an experiment must be reproduced, the same furnace and settings should be used. The furnace temperature settings and the position of the sample inside of the furnace tube had been deliberately calibrated to acquire a temperature gradient of 600~800°C, within which the δ, γ′, and γ″ phases may precipitate during the aging processes [19]. The temperature gradient during the aging process is stable without fluctuation, and the distribution of temperatures achieved at each monitored location is illustrated in Fig. 1(d). From Fig. 1(d), the experimentally obtained temperature gradient was within 605~825°C, which agreed well with our expectation. The aging process lasted 15 hours, followed by quenching into ice-water thereafter. The temperature diagram of heat treatment in the present work and the corresponding sample notations are summarized in Fig. 2. The alloy located adjacent to each thermocouple (as shown in Fig. 1(b) with a highlighted case for HT825) were sectioned individually for microstructure characterization to determine the effect of different aging temperatures. The samples were then surface polished using metallographic methods for SEM (scanning electron microscope) and EBSD (electron backscatter diffraction) characterizations. The SEM and EBSD were performed on an FEI Scios Dual-Beam system and were used for phase identification and grain morphology observation, respectively. In order to identify the $\gamma''/\gamma'$ particles with nanosize, higher resolution microstructure characterization was performed on an H-9500 E-TEM with an acceleration voltage of 300 kV. The TEM samples were twin-jet polished using an electropolisher under a voltage of 15 V at -30°C with a solution of 10 vol.% $HClO_4$ + 90 vol.% $CH_3OH$. Vickers microhardness testing was conducted with a Leco LM-800 indenter under a load of 100 gf and a dwell time of 10 s.

Figure 3(a) shows the effects of aging impact on microstructure hardness, which can indicate the strengthening effects. The hardness values demonstrate a parabolic relationship with the aging temperature gradient. Within the temperature range of 716~816°C, the hardness of the aged samples are higher than that in the wrought Inconel 718 (340 HV, AMS5662) [14], indicating the AM alloys could achieve higher strengthening effects when applied suitable heat treatment. The highest hardness is 477.5 $HV_{0.1}$ and occurs after aging at a temperature of 716°C. It is found that the temperatures above and below 716°C result in the reduction of hardness. The lowest hardness of 248.4 $HV_{0.1}$ is obtained at 605°C, which is lower than that in the as-built alloy (338 $HV_{0.1}$). The grain size of the as-built sample is 62.6±43.2 μm, while it increased to 128.1±75.2 μm after 1180°C-1h homogenization [18], and further increased to 169.1±86.09 μm in sample HT605 after aging without the presence of strengthening phases. Hence, the significantly increased grain size has a vital effect on the reduced hardness of the HT605 sample. Additionally, some other factors, like the evolution of residual stress after heat treatments may also contribute to the microhardness change, but more study is required for validation.

EBSD characterization was performed on each aged sample to observe the grain morphology. As presented in Figs. 3(b)-(i), coarse grains can be found to form in all aged samples. The average grain diameters measured by EBSD, in a range of 160~210 μm, are plotted as a function of the corresponding aging temperatures in Fig. 3(a). The grain size is shown to be independent of the aging temperature. Such observation elucidates that the chosen aging temperatures do not cause significant effects on the grain size and morphology, which has not yet been observed in the strengthening studies on Inconel 718. Additionally, the relatively large grain size achieved after heat treatment in this study has little contribution to the microhardness variation.

According to the microhardness testing results shown in Fig. 3(a), three representative samples were selected for further microstructure investigation to understand the structure-property relationships: (1) the sample HT605 having the lowest microhardness of 248.4 $HV_{0.1}$; (2) the sample HT716 having the highest microhardness of 477.5 $HV_{0.1}$; and (3) the sample HT825 owning the lowest microhardness of 332.2 $HV_{0.1}$ in the high-temperature gradient. As shown in the microstructure under SEM and TEM, Figs. 4(a)&(b), except a small amount of NbC carbides, no γ″/γ′ and δ precipitates can be observed in the sample HT605. In sample HT716 (Fig. 4(d)), the 716°C-aging causes a small amount of the δ phase to precipitate along grain boundaries. However, a large number of plate-shaped γ″ particles are observed in the TEM micrographs (Figs. 4(e)&(f)). These γ″ particles are very fine with a mean particle length of 13.8±4.2 nm through image analysis. The typical γ′ phase with spherical shape is not found to precipitate in sample HT716. This indicates that the precipitation of γ″ preceded the formation of γ′ in the current study. Therefore, the strengthening effect is dominated by γ″ with fine particle size. Figure 4(g) and its inset present significant precipitation of the δ phase along the grain boundaries in sample HT825. As pointed out by the red arrows in the inset of Fig. 4(g), some δ particles are observed not to have habit planes parallel to the grain boundaries, and their growth is usually accompanied by grain boundary migration. Hence, the corresponding grain boundaries have a jagged appearance, as discussed in Ref. [20]. Figure 4(h) shows the γ″ particles in sample HT825 are significantly coarsened with a mean particle length of 303.9±183.2 nm. Similarly to sample HT716, the γ′ phase is not observable in sample HT825, which suggests the precipitation of γ′ is still surpassed by γ″ during the aging at 825°C.

Besides the γ″ and δ phases, NbC carbides were found in samples HT716 and HT825 with a small phase fraction (less than 0.1%) in each sample, and thus would not influence the phase transformations and materials properties remarkably. In addition, some small black particles were found in the aged samples, which might be the (Nb,Ti)(C,N) carbonitrides or the oxide inclusions [21].

It is obvious that the different microhardness among the aged samples is due to the various phase transformation behaviors. According to Oblak et al. [22], the coherency strengthening stems from the γ″ phase is the primary mechanism for the hardening effects in Inconel 718. The coarsening of the γ″ precipitates can lead to the coherency loss and can weaken the hardening effects [23,24]. The critical mean particle length of the γ″ for coherency loss was reported to be 80~120 nm [17,23–25]. Consequently, it can be inferred that the very fine γ″ particles with a mean particle length of 13.4 nm obtained in 716°C-aging primarily contribute to the highest microhardness, since their particle size is smaller than the critical size, and these particles have high coherency with the γ matrix [24]. Correspondingly, the microhardness reduction in sample HT825 is mainly attributed to the coherency loss of γ″ [23,24], which is owing to the precipitates coarsening. Besides, the large amount of δ phase formed in sample HT825 is also considered to reduce the microhardness, because the δ precipitates have incoherent phase boundaries with the γ phase, which degrades the hardening effect in the alloy. However, in sample HT605, it is the absence of the strengthening phases that lead to the dramatically reduced microhardness, and this result suggests that the aging process at 605°C for 15 hours is insufficient for the precipitation hardening in AM Inconel 718.

Overall, the high-throughput experimentation coupling AM technique with GTHT can serve as an effective tool with flexibility to accelerate the post-heat treatment optimization and new alloy development for advanced manufacturing, especially for AM components. This method provides a new pathway to identify the process-structure properties for AM components with an expedited mode. The study on the Inconel 718 alloy has demonstrated that such a HT method would facilitate microstructure engineering. This study reveals that grain size is independent of the aging temperature between 605 and 825°C during isochronal aging. Aging at 716°C for 15 hours is found to generate the highest microhardness in this study, which is superior to the hardness of wrought materials. The aging temperature of 716°C with optimal microstructure is close to the widely used values of 718 or 720°C in the heat treatment of Inconel 718 [10,14], proving the validity in microstructure engineering of the proposed HT method. The hardness reduction in the high-temperature range is caused by the coarsening of γ″ particles. The precipitation of the δ phase can also result in the reduction of hardness. At low temperature around 605°C, the hardness reduction is due to the absence of the γ″ phase. The precipitation of the γ″ phase precedes the formation of the γ′ phase within the investigated temperature range.

## Acknowledgments

The authors thank the National Aeronautics and Space Administration (NASA) for the financial support under STTR Contract Number 80NSSC18C0214, N.S. and W.X. appreciate the support from NASA Space Technology Research Fellowship (NSTRF) Grant 80NSSC19K1142. We also appreciate the assistance by Ms. Yinxuan Li in sample preparation through the summer internship program supported by Mascaro Center for Sustainable Innovation at the University of Pittsburgh.

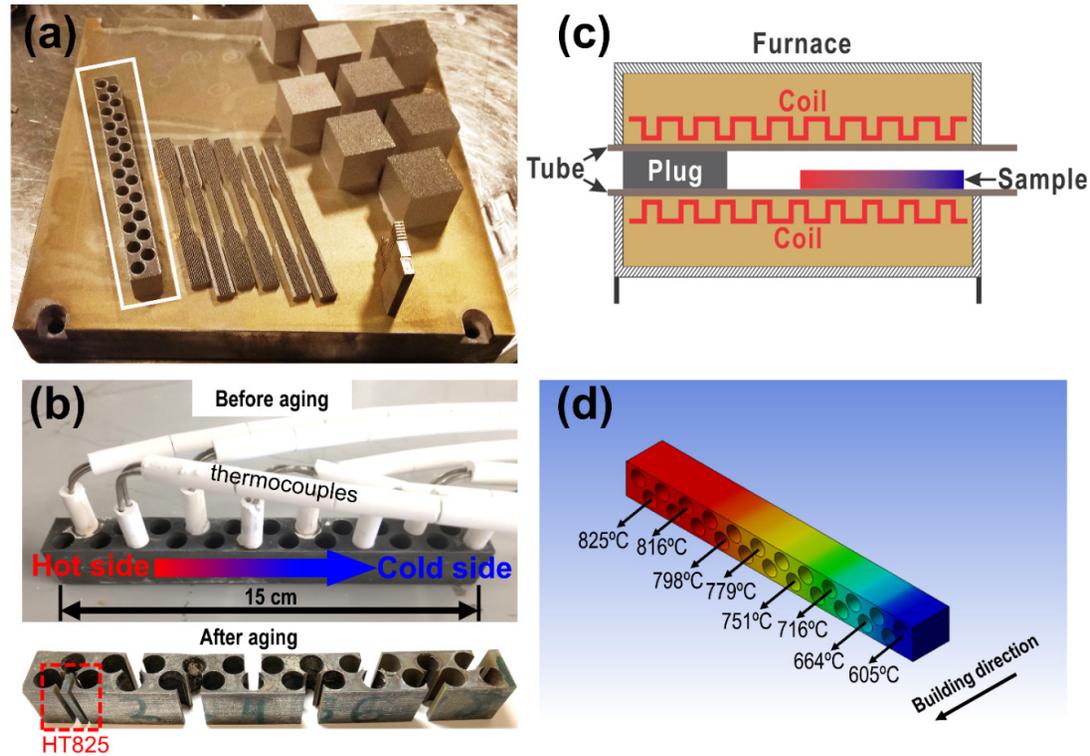

Figure 1. Setup of high-throughput experiments. (a) Inconel 718 build printed by LPBF; (b) setup of temperature record and illustration of sample cutting for microstructure characterization (taking sample HT825 as an example; the sample notations and the corresponding heat treatment are shown in Fig. 2); (c) setup of the furnace for the high-throughput experiment; (d) experimental temperature distribution inside the bar-sample.

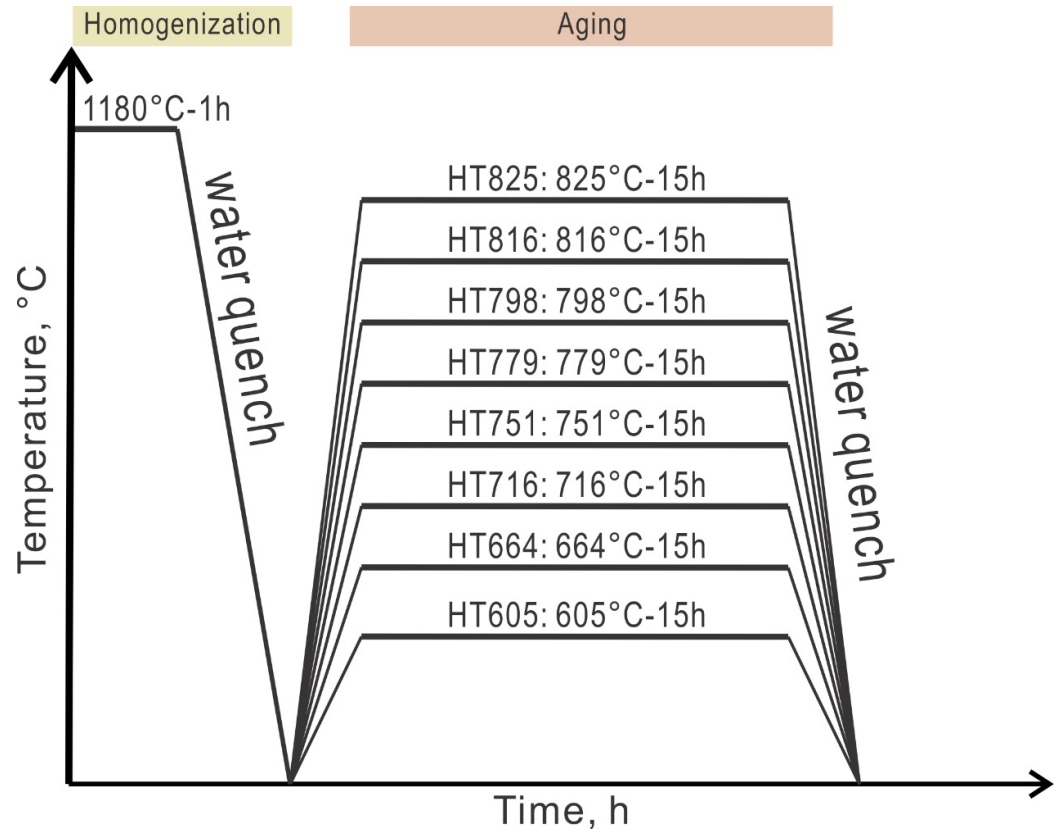

Figure 2. Temperature diagram of heat treatment with the corresponding sample notations used in this work.

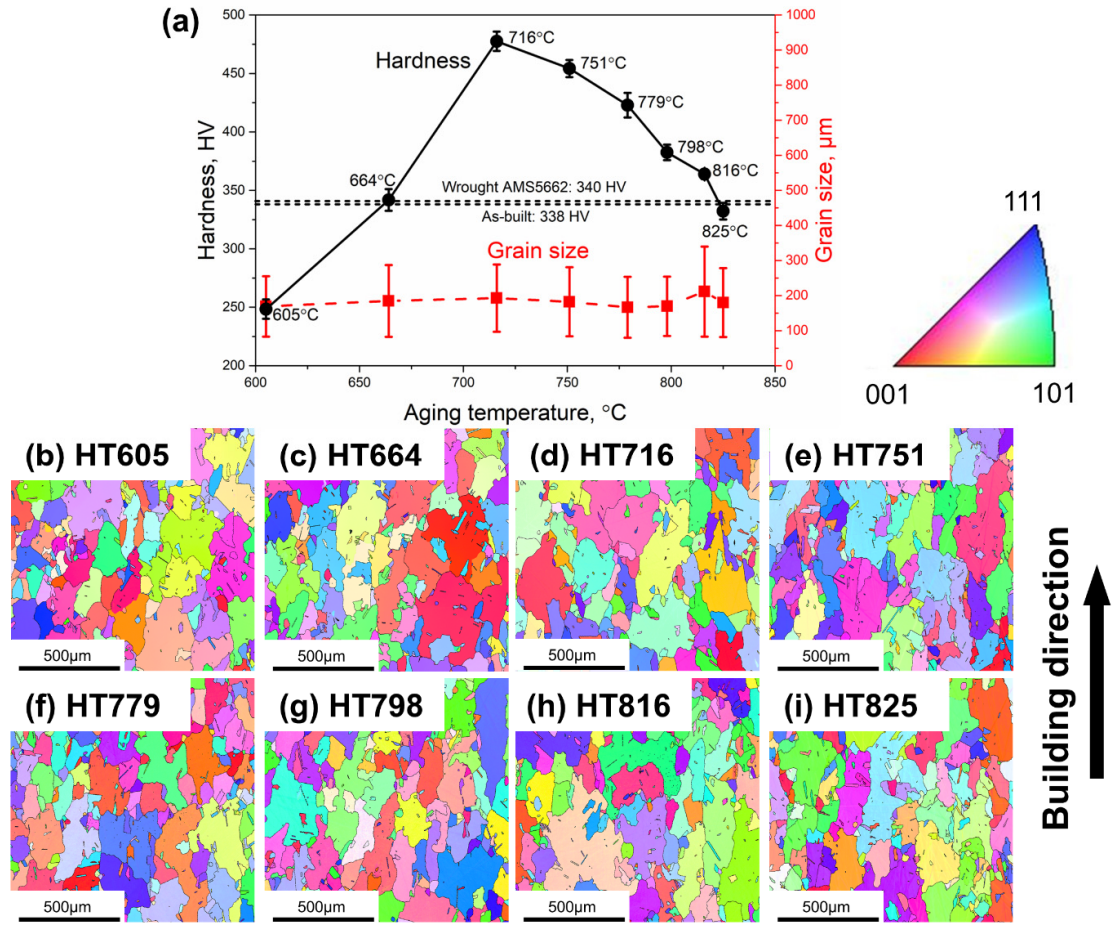

Figure 3. (a) Results of Vickers microhardness and average grain size measurements. IPFs of the aged samples with (b) HT605; (c) HT664; (d) HT716; (e) HT751; (f) HT779; (g) HT798; (h) HT816; (i) HT825.

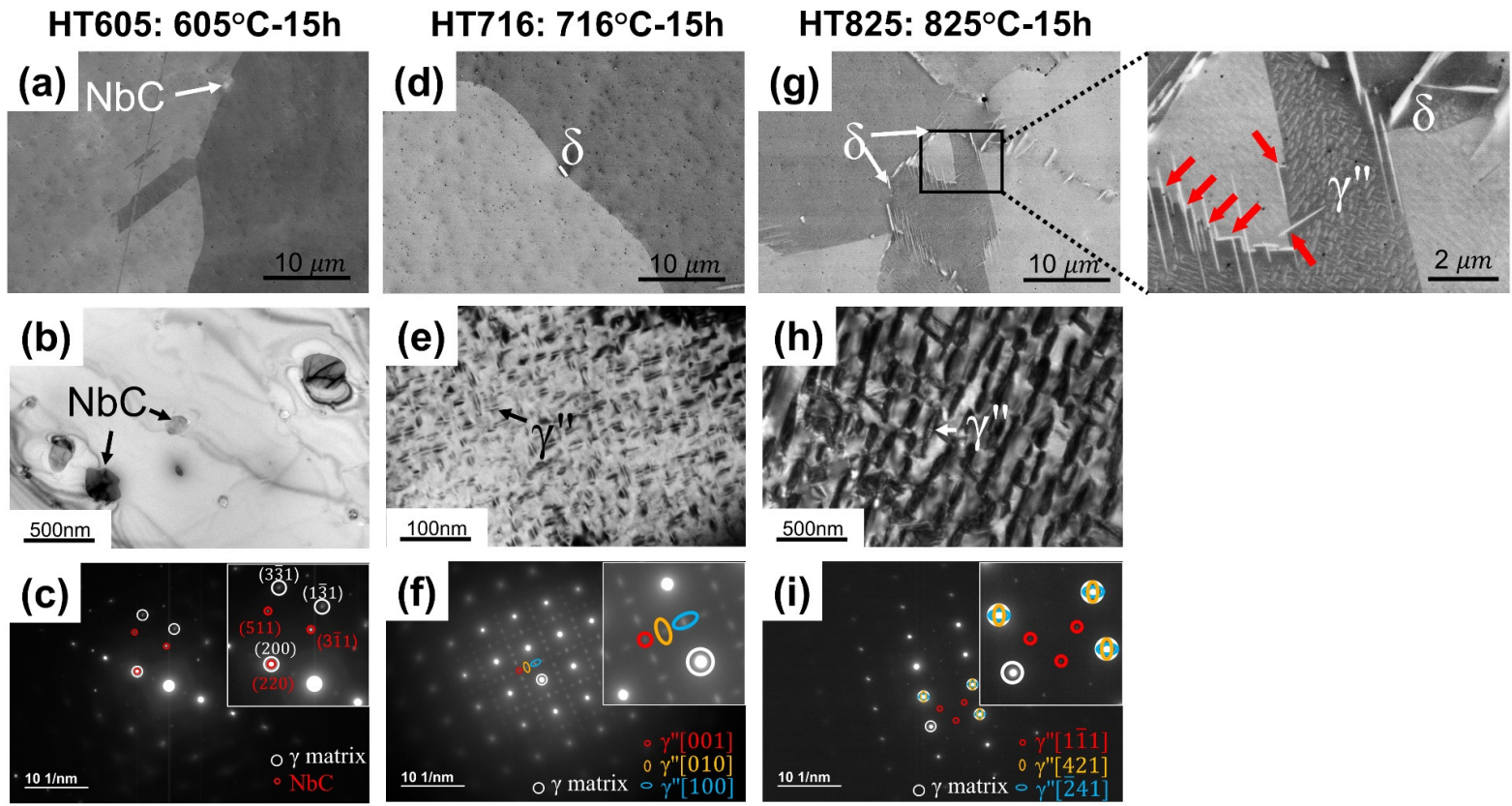

Figure 4. Microstructures of HT605 characterized by (a) SEM-BSE; (b) bright-field TEM; (c) selected-area-electron-diffraction (SAED). Microstructures of HT716 characterized by (d) SEM-BSE; (e) bright-field TEM; (f) SAED. Microstructures of HT825 characterized by (g) SEM-BSE; (h) bright-field TEM; (i) SAED. The different γ″ variants in (f) and (i) are shown by different colors, and the corresponding zone axes are indicated; the detailed indexing results are shown in Supplemental Information.

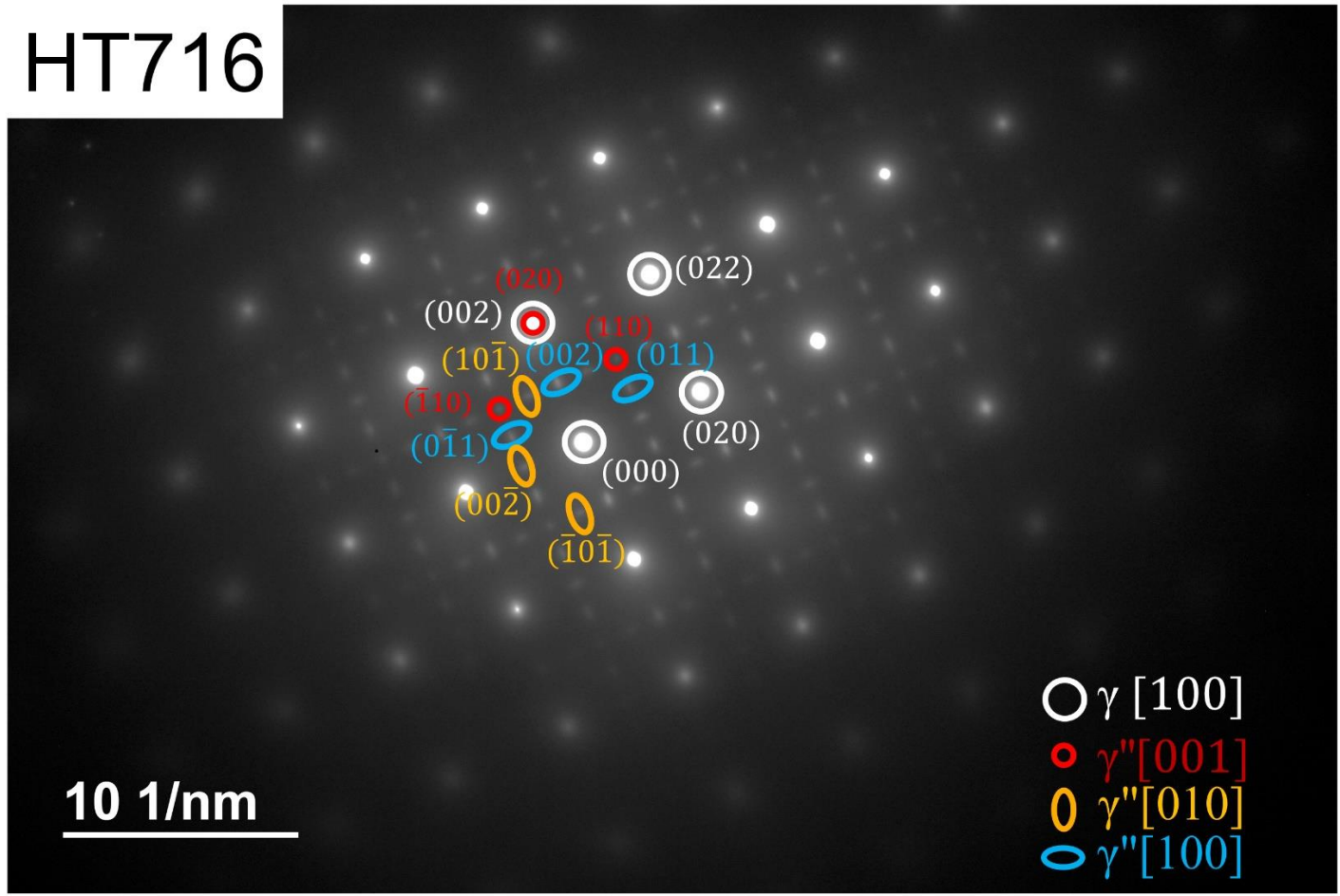

Figure S1. Diffraction pattern indexing for sample HT716. The different γ″ variants are shown by different colors, the corresponding zone axes are indicated at the right bottom corner.

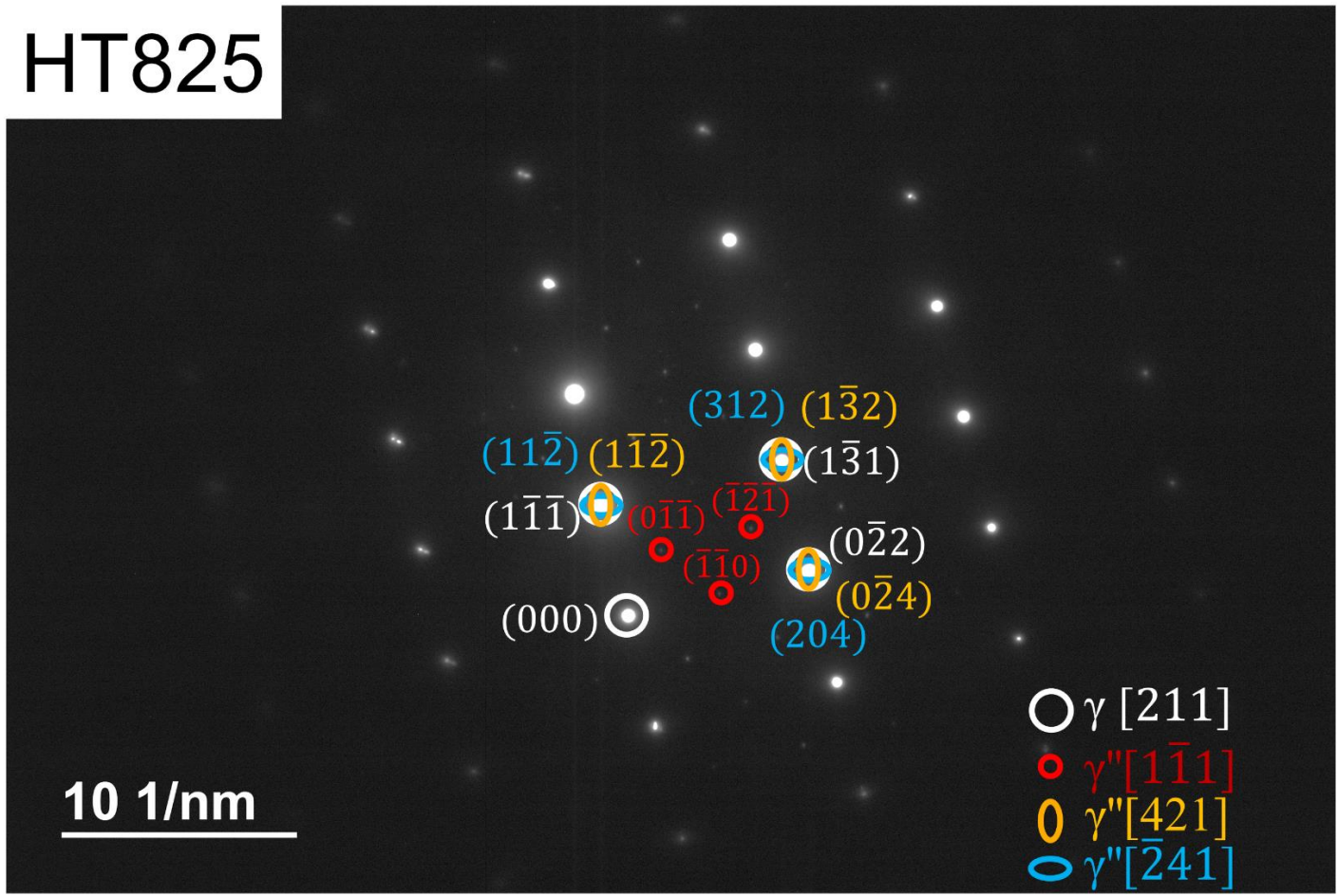

Figure S2. Diffraction pattern indexing for sample HT825. The different γ″ variants are shown by different colors, the corresponding zone axes are indicated at the right bottom corner.